\documentstyle[aps,prl,twocolumn,epsfig]{revtex}
\begin{document}

\title{Breached Pairing Superfluidity at Finite Temperature and Density}
\author{Jinfeng Liao and Pengfei Zhuang\\
       {\small Physics Department,Tsinghua University,Beijing 100084 }}

\maketitle

\begin{abstract}
\small A general analysis on Fermion pairing at finite temperature
and density between different species with mismatched Fermi
surfaces is presented. Very different from the temperature effect
of BCS phase, the recently found breached pairing phase resulted
from density difference of the two species lies in a region with
calabash-like shape in the $T-\mu$ plane, and the most probable
temperature for the new phase's creation is finite but not zero.
\end{abstract}


The Fermion pairing between different species with mismatched
Fermi surfaces, which has already been discussed many years ago in
investigating superconducting metal or B state of liquid $^3 He$
in a magnetic field\cite{loff,leggett}, prompted recently new
interest in both theoretical and experimental studies. In high
energy physics the strong interacting matter at high baryon
densities described by Quantum Chromodynamics (QCD) is a color
superconducting phase, in which quarks of different colors pair
with each other due to attraction in color anti-triplet
channel\cite{csc,sh}. The corresponding physical systems may be
found in compact stars\cite{glendenning}. In non-relativistic case
the study is related to the challenging goal of observing the BCS
transition in trapped Fermionic atoms\cite{lw1,wy,lw2,glw}, the
electrons in solid distributed in two different
bands\cite{lw1,glw}, and the neutron-proton pairing in isospin
asymmetric nuclear matter\cite{lombardo}.

When the two Fermions have the same Fermi surface, the pairing
phenomena is the well-known BCS mechanism\cite{bcs}: At low
temperature but high density, Fermionic matter with a sharp and
high Fermi surface is unstable under attractive interaction(even
very weak), a BCS paired state is favored instead. However, under
some physical constraints the two Fermions may come from different
Fermi surfaces. In atom traps, there can be two different
hyperfine states of the same atom($^6 Li$ or $^{40} K$) which
serve as two attracting species of fermions, and even mixture of
two different Fermionic atoms could be
realized\cite{lw1,wy,lw2,glw}. In QCD matter at moderate baryon
density (below the critical Color-Flavor-Locking density) which is
considered to be relevant for understanding neutron stars, the
mass difference between light quarks (u and d) and strange quark
can not be neglected, and therefore they have different Fermi
momenta\cite{misquark}. When chiral symmetry restoration at finite
density is considered\cite{chiral}, the mass difference becomes
more significant, and the mismatch will be enhanced. The mismatch
happens even for the pairing between the light quarks when charge
neutrality is taken into account\cite{neutrality}.

A well-known pairing mechanism of Fermions from mismatched Fermi
surfaces is the LOFF state\cite{loff}. Different from a BCS pair
which has zero total momentum, a LOFF pair has finite total
momentum. An analogue of this idea in dense QCD matter is the
crystalline color superconductivity\cite{crystaline}. Recently a
spatially uniform mechanism is proposed\cite{lw1} which leads to a
breached pairing superfluidity (BP)\cite{glw} with coexistence of
pairing state at lower and higher Fermi surfaces and particle
state in between, when the densities of the two species differ
from each other. This new phase is argued to be more favored than
the LOFF phase for certain region of parameters.

It is well-known that with increasing temperature of the system
the superfluidity gap in a BCS state drops down monotonously and
the region of superfluidity phase in the temperature-density plane
is reduced monotonously too\cite{temp}. What is the temperature
effect of a system in a BP state and what is the difference from
that in a BCS state? In this letter, we present a general analysis
on spatially uniform pairing state with mismatched Fermi surfaces
at finite temperature and density. We will discuss the stability
of BP state induced by mass difference of the two species, and
calculate the temperature behavior of the superfluidity gap and
the phase diagram for a system with two species of different
densities.

We start with a system containing two species of fermions
represented by $a$ and $b$. The interaction can be modelled by a
four-fermion point coupling, which is appropriate for both trapped
Fermionic atoms and dense quark systems\cite{csc,lw1,wy,lw2,glw}.
Since our purpose is a general analysis for pairing phenomena, we
neglect inner structures like spin, isospin, flavor and color,
which are important and bring much abundance while are not central
for pairing. We write down the following Hamiltonian,
\begin{eqnarray}
\label{hamiltonian} \hat{\cal H}= \frac{1}{V} \sum_{\vec
p}(\epsilon_{p}^a \hat{a}^{\dagger}_{\vec p} \hat{a}_{\vec
p}+\epsilon_{p}^b \hat{b}^{\dagger}_{\vec p} \hat{b}_{\vec p}) -
\frac{g}{V^2} \sum_{\vec p , \vec q} \hat{a}^{\dagger}_{\vec p}
\hat{b}^{\dagger}_{- \vec p} \hat{b}_{ - \vec q} \hat{a}_{ \vec
q}\ ,
\end{eqnarray}
where ${\bf p}$ and ${\bf q}$ are momenta of the species, $\hat a,
\hat b, \hat a^\dagger$ and $\hat b^\dagger$ are the annihilation
and creation operators, the coupling constant $g$ is positive to
keep the interaction attractive, $V$ is the system volume and in
continuous limit we simply replace $\frac{1}{V} \sum_{\vec p}$
with $\int \frac{d^3 {\bf p}}{(2\pi)^3}$, and the effective
particle energies $\epsilon_p^{a,b}$ are
$\sqrt{p^2+m_{a,b}^2}-\mu_{a,b}$ in relativistic case and
$\frac{p^2}{2m_{a,b}}-\mu_{a,b}$ in non-relativistic case. The
Fermi momenta $p_F^{a,b}$ determined by $\epsilon_p^{a,b} = 0$ are
controlled by the particle masses $m_{a,b}$ and the chemical
potentials $\mu_{a,b}$. Whether $\mu_{a,b}$ can be used as free
parameters depends on the physical system we discuss. When the
densities of the species are restricted by some physical
constraint like fixed overall particle density or fixed relative
particle density, $\mu_{a,b}$ are not fully free and should be
adjusted to satisfy the constraint. In fact, it is the density
constraint which brings nontrivial BP state, as pointed out in
\cite{lw1,wy,lw2,glw} and will be clearly shown below.

We introduce in the light of Mean-Field an order parameter
$\Delta= \frac{g}{V} \sum_{\vec p}<\hat{a}^{\dagger}_{\vec p}
\hat{b}^{\dagger}_{- \vec p}>$ and make it real by proper choice
of phase factors of the creation operators, which allows us to
diagonalize the Hamiltonian into
\begin{eqnarray} \label{diag}
\hat{\cal H}_{diag}=\frac{1}{V} \sum_{\vec p}(E_p^A
\hat{A}^{\dagger}_{\vec p} \hat{A}_{\vec p} + E_p^B
\hat{B}^{\dagger}_{\vec p} \hat{B}_{\vec p} + \epsilon_p^b ) +
\frac{\Delta^2}{g}
\end{eqnarray}
after a Bogliubov transformation from Fermions $a$ and $b$ into
quasi-Fermions $A$ and $B$ with annihilation and creation
operators $\hat A, \hat B, \hat A^\dagger, \hat B^\dagger$ and
quasi energies
\begin{equation} \label{quasi}
 E_p^{A,B} = \epsilon_p^- \pm
\sqrt{\epsilon_p^{+2}+\Delta^2}
\end{equation}
with
\begin{equation}\label{energy}
\epsilon_p^{\pm} = \frac{\epsilon_p^a \pm \epsilon_p^b}{2}.
\end{equation}

\begin{figure}[ht]
\hspace{+0cm} \centerline{\epsfxsize=6cm\epsffile{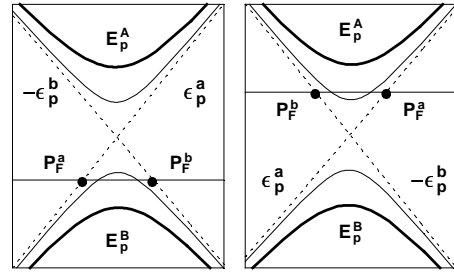}}
\caption{\it Dispersions of particles (dashed lines) and
quasi-particles (solid lines). The thick and thin solid lines
correspond to $\Delta<\Delta_c$ and $\Delta>\Delta_c$,
respectively. The left panel is for $E_p^B =0$, and the right
panel is for $E_p^A = 0$.}\label{fig1}
\end{figure}

The dispersions for particles $a,b$ and quasi-particles $A,B$ in
relativistic case are shown in Fig.(\ref{fig1}). With the interest
in mismatched Fermi momenta induced by mass difference and density
difference, a remarkable feature is that $E_p^{A,B}$ can cross
zero between the particle Fermi momenta $p_F^{a,b}$. A simple
algebra calculation shows that the condition for $E_p^{A(B)}$ to
cross zero is the constraint $\Delta < \Delta_c$ for the gap. In
relativistic case with the approximation of $p_F^{a,b} >>
m_{a,b}$, the critical gap is
\begin{equation} \label{dcr}
\Delta_c = \frac{\sqrt{p_F^a \cdot p_F^b}}{2 \sqrt{\mu_a \mu_b}}
|p_F^a - p_F^b|\ ,
\end{equation}
in non-relativistic case one has exactly
\begin{equation} \label{dcnr}
\Delta_c = \frac{p_F^a+p_F^b}{4 \sqrt{m_a m_b}} |p_F^a - p_F^b|\ .
\end{equation}
For convenience, we introduce momentum interval $I \equiv \{
\vec{p} | p_- < |\vec p| < p_+ \}$ with $p_{\pm}$ the solutions of
$E_p^{A(B)}=0$.

Whether $E_p^{A(B)}$ crosses zero along the momentum axis plays a
crucial role in pairing with mismatched Fermi surfaces.
$E_p^{A(B)}=0$ means directly gapless excitation of
quasi-particles in a superfluidity phase observed in
\cite{sh,lw1,wy,lw2,glw,gapless}. We see from Fig.(\ref{fig1})
large mismatch and small gap leave space for gapless phenomena. If
there is no mismatch, there is no place for gapless excitation.

In order to understand the coexistence of pairing state and
particle state when the gapless excitation region $I$ is not
empty, we write down the ground state of the system in terms of
the zero quasi-particle state $|0^{A(B)}_p>$ and one
quasi-particle state $|1^{A(B)}_p>$ at zero temperature,

%%%%%%%%%%%%%%%%%%%%%%%%%%%%%%%%%%%%%%%%%%%%%%%%%%%%%%%%%%%%%%%%%%%%%%%%
%%%%%%%%%%%%%%%%%%%%%%%%%%%%%%%%%%%%%%%%%%%%%%%%%%%%%%%%%%%%%%%%%%%%%%%%

\begin{eqnarray} \label{ground}
|g>=\prod_{\vec{p}} && {\bigg(} \theta(E_p^A)\theta(-E_p^B)|
0^A_p, 1^B_p > \nonumber \\
&& + \theta(-E_p^A) |1^A_p, 1^B_p> + \theta(E_p^B) |0^A_p, 0^B_p>
{\bigg )}
\end{eqnarray}
where $\theta(x)$ is a step function. With the help of the
Bogliubov transformation the ground state can be expressed in the
space of the original Fermions $a$ and $b$,
\begin{eqnarray} \label{occupy}
|g>= && \prod_{\vec{p}\notin I} {\big(}
cos\theta_{p}-sin\theta_{p} \hat{a}^{\dagger}_{\vec p}
\hat{b}^{\dagger}_{-\vec p} {\big)}|0^a_p,0^b_p> \times\nonumber
\\
&& \prod_{\vec{p}\in I} {\big ( } \theta(p_F^a -p_F^b)
\hat{a}^{\dagger}_{\vec p} + \theta(p_F^b - p_F^a)
\hat{b}^{\dagger}_{-\vec p} {\big )} |0^a_p,0^b_p>\ .
\end{eqnarray}
The above state tells us clearly the breached pairing phenomena.
If neither $E_p^A$ nor $E_p^B$ crosses zero, namely without
gapless excitation, the species $a$ and $b$ are symmetrically and
partially occupied and paired, the system is in a BCS pattern. If
one can find roots $p_\pm$ from $E_p^{A(B)} = 0$, there exists a
momentum interval $I$ in which one specie is fully occupied and
the other is fully empty. The pairing occurs only in the region
around the lower Fermi surface and the region around the higher
Fermi surface, the two regions are separated by the gapless
excitation region $I$. This means a new phase --- the breached
pairing superfluidity proposed in \cite{glw}. The observed gapless
phenomena \cite{sh,lw1,wy,lw2,gapless} all fall into the BP phase,
which is characterized by two universal features, pairing breached
by single occupation and simultaneous gapless and gaped
components.

We now turn to thermodynamics. From the Hamiltonian (\ref{diag})
in quasi-Fermion representation the grand potential is easily
expressed as
\begin{eqnarray} \label{omega}
\Omega(T,\mu_a,\mu_b,\Delta)&=& \frac{\Delta^2}{g} - \frac{1}{V}
\sum_{\vec p} {\bigg [ }   {\big ( } \frac{E_p^A}{2} + T Ln(1+e^{-
E_p^A /T}) {\big ) } \nonumber \\
 &+& {\big ( } \frac{E_p^B}{2} + T Ln(1+e^{-
E_p^A / T}) {\big ) } - {\epsilon}_p^+ {\bigg ]}\ .
\end{eqnarray}
With given $\Omega$ as a function of temperature, chemical
potentials and gap parameter, all other thermodynamical quantities
can be immediately obtained and its minimum gives the gap equation
for solving $\Delta$
\begin{equation} \label{gap}
\Delta \cdot {\bigg [}  \frac{2}{g}-\frac{1}{V} \sum_{\vec p}
\frac{f(E_p^B)-f(E_p^A)}{\sqrt{\epsilon_p^{+ 2}+\Delta^2}} {\bigg
]} =0
\end{equation}
with the Fermi-Dirac distribution $f(x)=\frac{1}{e^{x/T}+1}$.

To do numerical calculation and present quantitative results we
consider in the following only relativistic case. The calculation
can straightforwardly be extended to non-relativistic case. Since
the four-Fermion interaction is not renormalizable, one must use a
momentum cut-off $\Lambda$ to avoid divergence. We choose $\Lambda
= 600$ MeV in our numerical calculation. The mismatch of the Fermi
surfaces can be induced by either mass difference or density
difference of the two species. We now solve the gap equation
(\ref{gap}) at finite temperature in the two cases.

{\it Mass Difference ---} We consider a system with
$\mu_{a,b}=\mu$ but $m_a \ne m_b$ to study the effect of mass
difference. For given $m_a, m_b, \mu$ and $T$ one can define two
couplings $g_{c1} < g_{c2}$, corresponding to the critical
conditions to have a BP state and a BCS state, respectively,
\begin{eqnarray} \label{coupling}
g_{c1}&&=  {\bigg [}  2 {\bigg /} \frac{1}{V} \sum_{\vec p}
\frac{f(E_p^B)-f(E_p^A)}{\sqrt{\epsilon_p^{+ 2}+\Delta^2}} {\bigg
]} {\bigg |}_{\Delta \to \Delta_c} \quad ,
\nonumber \\
g_{c2}&&=  {\bigg [}  2 {\bigg /} \frac{1}{V} \sum_{\vec p}
\frac{f(E_p^B)-f(E_p^A)}{\sqrt{\epsilon_p^{+ 2}+\Delta^2}} {\bigg
]} {\bigg |}_{\Delta \to 0} \quad .
\end{eqnarray}

%%%%%%%%%%%%%%%%%%%%%%%%%%%%%%%%%%%%%%%%%%%%%%%%%%%%%%%%%%%%%%%%%%%%%%%%
\begin{figure}[ht]
\hspace{+0cm} \centerline{\epsfxsize=6cm\epsffile{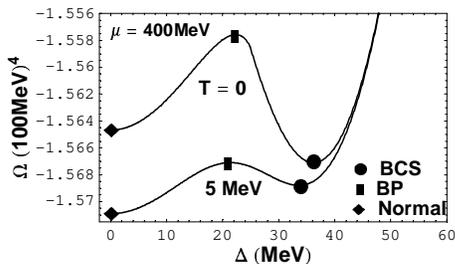}}
\caption{\it The potential $\Omega$ as a function of $\Delta$ at
given $T,\mu$ with $m_a=5 MeV, m_b=200 MeV$, and $g=50\
GeV^{-2}$.}\label{fig2}
\end{figure}
%%%%%%%%%%%%%%%%%%%%%%%%%%%%%%%%%%%%%%%%%%%%%%%%%%%%%%%%%%%%%%%%%%%%%%%%

\noindent For $g<g_{c1}$ the only solution of (\ref{gap}) is
$\Delta=0$ which means a normal phase without pairing. For
$g>g_{c2}$ there is a stable BCS solution and $\Delta = 0$ becomes
unstable. When $g$ is in the region $g_{c1}<g<g_{c2}$, there are
three solutions of (\ref{gap}): a normal solution $\Delta=0$, a
BCS solution $\Delta>\Delta_c$  and a BP solution $\Delta <
\Delta_c$. However, only the BCS solution is the global minimum of
$\Omega$, the BP solution is unstable because it doesn't satisfy
the stable condition $\partial^2 \Omega/\partial\Delta^2 > 0$, and
the normal solution is metastable. This is shown clearly in
Fig.(\ref{fig2}). When the temperature increases, the normal
solution becomes stable and the BCS solution changes to be
metastable. In any case the BP solution can not be stable. When
$T$ exceeds some critical value, the unstable BP state even
disappears. Therefore, the stability analysis rules out the
possibility to create a BP phase by mass difference.

{\it Density Difference ---} From the thermodynamic potential
(\ref{omega}) it is easy to obtain the particle densities
\begin{eqnarray} \label{dendity}
n_a = {1\over 2V}\sum_{\vec p}{\bigg [}&&
\left(1+{\epsilon_p^+\over
\sqrt{\epsilon_p^{+2}+\Delta^2}}\right)f(E_p^A)\nonumber\\
&&+\left(1-{\epsilon_p^+\over
\sqrt{\epsilon_p^{+2}+\Delta^2}}\right)f(E_p^B){\bigg ]}\ ,\nonumber\\
n_b = {1\over 2V}\sum_{\vec p}{\bigg [}&&
-\left(1-{\epsilon_p^+\over
\sqrt{\epsilon_p^{+2}+\Delta^2}}\right)f(E_p^A)\nonumber\\
&&-\left(1+{\epsilon_p^+\over
\sqrt{\epsilon_p^{+2}+\Delta^2}}\right)f(E_p^B)+2{\bigg ]}\ .
\end{eqnarray}

We consider a system with fixed relative density
\begin{equation}
\label{lambda} {n_b\over n_a} = \lambda\ .
\end{equation}
To satisfy this density constraint, namely a charge conservation
such as electronic charge with $a,b$ carrying charge number
$-\lambda,+1$, respectively, the chemical potentials should be
adjusted to be $\mu_a=\mu-\lambda \delta \mu$ and $\mu_b=\mu +
\delta \mu$, where $\mu$ corresponds to the total number density
since $n_a + n_b = -\partial \Omega/\partial\mu$. Unlike the case
of mass difference where the chemical potentials are free and
$\Delta$ is purely determined by the gap equation, here only $\mu$
is free and $\Delta$ and $\delta \mu$ are coupled to each other
through (\ref{gap}) and (\ref{lambda}).

\begin{figure}[ht]
\hspace{+0cm} \centerline{\epsfxsize=6cm\epsffile{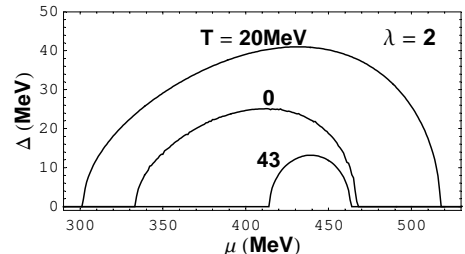}}
\caption{\it The gap parameter as a function of $\mu$ at fixed
relative density $\lambda =2$ for different
temperature.}\label{fig3}
\end{figure}

For $\lambda = 1$ the system can be in a BCS state, while for
$\lambda \ne 1$ there will be some particles left after the
pairing, the system must be in a BP state if pairing occurs. We
first calculate $\Delta$ as a function of $\mu$ for $\lambda = 2$
at given temperature. We choose the parameters $m_a = m_b = 5$
MeV, and $g=80\ GeV^{-2}$ in the following numerical calculation.
Introducing mass difference in calculation brings only
quantitative changes. The value of $g$ needed to produce a BP
decreases fast with increasing momentum cut-off $\Lambda$, and
taking into account all inner degrees of freedom like spin,
isospin, color and flavor will reduce the coupling by order of
magnitude. From Fig.(\ref{fig3}) a BP phase exists in a relatively
narrow region, compared with a BCS state. The explanation for the
existence of the higher critical chemical potential is that when
$\mu$ is high enough the mismatch of the two Fermi surfaces is too
large for the two species to pair at given coupling.

In a BCS state it is well-known that the gap and the region of
superfluidity decrease monotonously with increasing temperature.
However, the case in a BP state is very different. With increasing
temperature, the gap firstly goes up and then drops down, and
correspondingly the region firstly expands and then contracts. The
temperature related to the maximum gap and the maximum region is a
finite value but not zero. We can understand this feature in the
following way. When the two Fermi surfaces coincide, the
temperature deforms and lowers the Fermi surface and then the gap
decreases. In the case of mismatch, the two different sharp Fermi
surfaces are deformed and lowered by the temperature on one hand,
but the expansion of the two distributions make the two species
much closer in the phase space and then much easier to pair on the
other hand. The former temperature effect reduces the gap but the
latter effect favors the gap. The competition between the former
and the latter effects controls the temperature dependence of the
amplitude and the region of superfluidity.

\begin{figure}[ht]
\hspace{+0cm} \centerline{\epsfxsize=6cm\epsffile{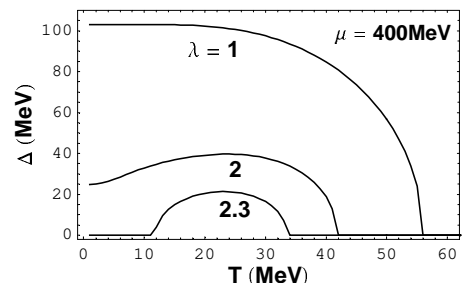}}
\caption{\it The gap parameter as a function of $T$ at fixed
chemical potential for different relative density.}\label{fig4}
\end{figure}

The behavior of the gap as a function of temperature at fixed
chemical potential for different relative density is shown in
Fig.(\ref{fig4}). At $\lambda = 1$, the gap drops down
monotonously due to disorder brought in by increasing temperature,
a standard characteristic of BCS phase. For $\lambda\ne 1$, we see
again the amazing temperature behavior: The gap first
increases to a maximum value and then reduces rapidly to zero with
increasing temperature. While the density difference $\lambda \ne
1$ is the prerequisite for a BP phase, the amplitude
and the region of the gap decreases with increasing $\lambda$.

\begin{figure}[ht]
\hspace{+0cm} \centerline{\epsfxsize=6cm\epsffile{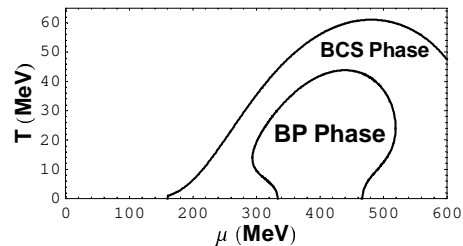}}
\caption{\it The BCS ($\lambda = 1$) and BP ($\lambda = 2$) phase
transition lines in $T-\mu$ plane.}\label{fig5}
\end{figure}

We further give in Fig.(\ref{fig5}) the phase diagram in $T-\mu$
plane for the BP phase with $\lambda = 2$ and compare it with the
BCS phase diagram with $\lambda = 1$. The calabash-like phase
transition line from normal state to BP state manifests the
competition between the two opposite temperature effects on the
mismatched Fermi surfaces, discussed above. The most probable
temperature for producing a BP state is not zero but a finite
value.

In summary, we have presented a general analysis on Fermion
pairing between different species with mismatched Fermi surfaces
under the assumption of spatial uniform, and investigated
especially the temperature behavior of the new BP phase. From the
stability analysis the mass difference can not create a stable BP
state. In the case of fixed relative density $n_b/n_a \ne 1$ the
temperature effect not only deforms and reduces the mismatched
Fermi surfaces which leads to the usual suppression of the gap,
but also make the overlap region of the two species even wider
which favors the condensate. The competition of the two opposite
temperature effects results in a calabash-like phase transition
line from normal to BP states. The temperature corresponding to
the largest gap and the largest region of superfluidity is finite
but not zero. Due to this novel temperature effect, we think that
trapping different Fermionic atoms with different densities at
controlled finite temperature may be a feasible way to obtain a BP
phase.

We are grateful to Dr. M. Huang for her stimulating discussions.
The work was supported in part by the NSFC under contract numbers
19925519, 10135030 and 10105005.

%%%%%%%%%%%%%%%%%%%%%%%%%%%%%%%%%%%%%%%%%%%%%%%%%%%%%%%%%%%%%%%%%%%%%%%
%%%%%%%%%%%%%%%%%%%%%%%%%%%%%%%%%%%%%%%%%%%%%%%%%%%%%%%%%%%%%%%%%%%%%%%%

%%%%%%%%%%%%%%%%%%%%%%%%%%%%%%%%%%%%%%%%%%%%%%%%%%%%%%%%%%%%%%%%%%%%%%%
%%%%%%%%%%%%%%%%%%%%%%%%%%%%%%%%%%%%%%%%%%%%%%%%%%%%%%%%%%%%%%%%%%%%%%%%

%%%%%%%%%%%%%%%%%%%%%%%%%%%%%%%%%%%%%%%%%%%%%%%%%%%%%%%%%%%%%%%%%%%%%%%
%%%%%%%%%%%%%%%%%%%%%%%%%%%%%%%%%%%%%%%%%%%%%%%%%%%%%%%%%%%%%%%%%%%%%%%%

\end{document}